\documentclass[prd,aps,nofootinbib,eqsecnum, superscriptaddress]{revtex4}

\usepackage{amsmath,amssymb,bm}
\usepackage[colorlinks]{hyperref}
\usepackage[all]{hypcap}
\usepackage{graphicx,psfrag,float,epsfig}
\usepackage{mathrsfs,wasysym}
\usepackage{epstopdf}
\usepackage{comment}
\usepackage{pbox}
\usepackage{array}
\usepackage{xspace}
\usepackage[usenames,dvipsnames]{color}
\usepackage[normalem]{ulem}

\newcommand{\nn}{\nonumber}
\newcommand{\beq}{\begin{equation}}
\newcommand{\eeq}{\end{equation}}
\newcommand{\bea}{\begin{eqnarray}}
\newcommand{\eea}{\end{eqnarray}}

\def\ba{ {\bm a} }
\def\bL{ {\bm L} }
\def\bn{ {\bm n} }

\def\br{ {\bm r} }
\def\bS{ {\bm S} }
\def\bv{ {\bm v} }
\def\bx{ {\bm x} }

\def\bSigma{ {\bm \Sigma} }

\def\lbracket{ { \big \{ \!\! \big  \{ } }
\def\rbracket{ {\big \} \!\! \big \} } }

\def\addICTP{ICTP South American Institute for Fundamental Research,\\ Rua Dr. Bento Teobaldo Ferraz 271, 01140-070 S\~ao Paulo, SP Brazil}

\def\addIFT{Instituto de F\'isica Te\'orica, Universidade Estadual Paulista,\\ Rua Dr. Bento Teobaldo Ferraz 271, 01140-070 S\~ao Paulo, SP Brazil}

\def\addUPitt{Pittsburgh Particle Physics Astrophysics and Cosmology Center, Department of Physics and Astronomy, University of Pittsburgh, Pittsburgh, PA 15260, USA}

\def\addJPL{Jet Propulsion Laboratory, California Institute of Technology, Pasadena, CA 91109, USA}

\def\addCaltech{Theoretical Astrophysics, Walter Burke Institute for Theoretical Physics, California Institute of Technology, Pasadena, CA 91125, USA}

\begin{document}

\title{Radiation reaction for spinning bodies in effective field theory II:~Spin-spin effects}

\author{Nat\'alia T.~Maia}
\affiliation{\addIFT}

\author{Chad R.~Galley}
\affiliation{\addJPL}
\affiliation{\addCaltech}

\author{Adam K.~Leibovich}
\affiliation{\addUPitt}

\author{Rafael A.~Porto}
\affiliation{\addIFT}
\affiliation{\addICTP}

\begin{abstract}
We compute the leading Post-Newtonian (PN) contributions at quadratic order in the spins to the radiation-reaction acceleration and spin evolution for binary systems, entering at four-and-a-half PN order. Our calculation includes the back-reaction from finite-size spin effects, which is presented for the first~time. 
The computation is carried out, from first principles, using the effective field theory framework for spinning extended objects. At this order, nonconservative effects in the spin-spin sector are independent of the  spin supplementary conditions. A non-trivial consistency check is performed by showing that the energy loss induced by the resulting radiation-reaction force is equivalent to the total emitted power in~the far zone. 
We find that, in contrast to the spin-orbit contributions (reported in a companion paper), the radiation reaction affects the evolution of the spin vectors once spin-spin effects are incorporated. 
\end{abstract}

\maketitle

\section{Introduction}

Motivated by the advent of gravitational wave astronomy \cite{Abbott:2016blz,2016htt,2016pea}, in this paper we continue to study spin effects on radiation reaction for the dynamical inspiral of binary compact objects.  In a companion paper \cite{paper1}, we reported the resulting radiation-reaction equations of motion to linear order in the spins, which appear at fourth Post-Newtonian (4PN) order, using the effective field theory (EFT) formalism. At this order conservative contributions are also known using more traditional methods \cite{blanchet} as well as the EFT approach \cite{nrgr,nrgrs}, encompassing spin-independent \cite{4pnjs,4pnjs2,4pndjs,4pndjs2,4pnbla,4pnbla2,4pndj,nrgr4pn1,nltail,nrgr4pn2,tailfoffa} (see also \cite{lambzero,comp}), and spin-dependent \cite{eih,3pnproc,comment, nrgrss,nrgrs2, nrgrso,Levi:2008nh,Perrodin:2010dy,Levi:2010zu,owen,buo1,buo2,damournloso,Schafer3pn,Schafer3pn2,schaferEFT,HergtEFT,hartung, bohennloso,luc12,steinhoffnnlo1,levinnlo1,levinnlo2,levinnlo3,equiv4pn,bohennloss}  terms.\footnote{The radiated power has been obtained to 3.5PN order also for spinning bodies \cite{srad,amps,bohennloss,luc13}, albeit not yet to 4PN other than the effects due to the next-to-leading order tail \cite{luc132} and gravitational wave absorption \cite{dis1,abspin}. The nonconservative effects of radiation reaction from non-spinning bodies were also computed~in~\cite{chadbr1, chadbr2} in the EFT approach.}
See~\cite{blanchet,Buoreview} for extensive reviews on the Post-Newtonian expansion, and~\cite{nrgrLH,rafric,iragrg,rafgrg,riccardocqg,review} for reviews on the EFT approach to the binary problem.\vskip 4pt

The purpose of the present work is to compute --from first principles-- the radiation-reaction acceleration and spin evolution for binary systems to 4.5PN order, quadratic in the spins. We achieve this using the EFT framework for spinning bodies~\cite{review} extended to nonconservative systems~\cite{chadgsf,chadbr1,chadprl,chadprl2,chadbr2}.  At 4.5PN order, the computations are independent of the spin supplementary condition. Some of these results were previously computed in \cite{Will2} (and \cite{RRADM} for the radiation reaction Hamiltonian) using different methodologies. However, we report --for the first time-- the effects of back-reaction due to the finite-size of the spinning bodies, computed from first principles.\footnote{See \cite{G3,G4,G5,GV} for related work relying on energy balance equations.}  We perform a consistency check similar to the one presented in~\cite{paper1}, that demonstrates the equivalence between the energy loss induced by the radiation reaction acceleration and the total emitted power in the far zone, up to total time derivatives.  
Contrary to what occurs in the spin-orbit sector \cite{Will1,paper1}, we find that spin-spin effects in the radiation reaction do affect the evolution of the spin vectors, leading to a radiation-reaction induced contribution to the full precession equation consistent with the findings in~\cite{Will2}, this time also including finite-size effects. Throughout this paper we use the same conventions as in \cite{paper1}, to which we also refer the reader for background discussion and references regarding details about some of the methods employed in this work.

\section{Radiation Reaction for Spinning Bodies}

\subsection{The Routhian approach and finite-size effects}

The conservative dynamics for a binary system of spinning bodies can be obtained from a Routhian,
\begin{align}
\label{eq:Routh}
	{\cal R} =  - \left( m \sqrt{u^2} + \frac{1}{2} \omega_\mu{}^{ab} S_{ab} u^\mu + \frac{1}{2m }R_{\nu\alpha\rho\sigma}S^{\rho\sigma}u^{\nu} S^{\alpha\beta} u_\beta + \cdots\right)\,,
\end{align}
with $S^{ab}$ the spin tensor in a locally Minkowski frame. The equations of motion follow from 
\begin{align}
	\frac{\delta }{\delta x^\mu} \int  {\cal R} ~ d\sigma = 0 ~, \qquad \frac{d
S^{ab}}{d\sigma} = \{ S^{ab}, {\cal R} \} \label{eq:eom},
\end{align}
and the spin algebra 
\begin{align}
	\{ S^{ab},S^{cd} \} = \eta^{ac} S^{bd} +\eta^{bd}S^{ac}-\eta^{ad} S^{bc}-\eta^{bc}S^{ad}\,.\label{eq:algebra}
\end{align}
The third term in \eqref{eq:Routh} in principle contributes to the spin-spin sector, however it can be shown to be subleading and therefore we will ignore it in what follows. 
In order to incorporate finite-size effects we need to augment the number of terms in \eqref{eq:Routh} beyond the {\it minimal coupling}. 
The first contribution appears already at leading order and it is a `self-induced' effect due to the rotation of the bodies. 
In the EFT framework, this effect is described by an additional term in the Routhian
\beq
\label{eq:finiteSize1}
\frac{C^{(a)}_{ES^2}}{2 m_a} \, \frac{E_{ab}}{\sqrt{u^2}} {S^a}_c S^{cb}\,,
\eeq
where $C^{(a)}_{ES^2}$ is a {\it Wilson coefficient} that encapsulates the short-distance properties of the object, and 
\beq
E_{ab} \equiv e_a^\mu e_b^\nu C_{\mu\alpha\nu \beta} u^\alpha u^\beta\,
\eeq
is the electric component of the Weyl tensor, $C_{\alpha\beta\gamma\rho}$, with $e^\mu_a$ a locally flat tetrad field. 
For the case of black holes we have $C_{ES^2} =1$ (hence the normalization in~\eqref{eq:finiteSize1}), while it is an order of magnitude larger for neutron stars. 
We refer the reader to \cite{review} for a more detailed exposition.

From the Routhian (including the term in~\eqref{eq:finiteSize1}) we can obtain the leading order spin-spin equations of motion in the conservative sector, both with ${\cal O}(\bS_1\bS_2)$ and ${\cal O}(\bS_a^2)$ terms, namely, the relative accelerations that enter at 2PN order,
\begin{equation}
\ba^{s_1s_2}_{\rm cons}=-\frac{3}{m\nu r^{4}}\left[  \bn\left(
\bS_{1}\cdot\bS_{2}\right)  -5\bn\left(  \bS%
_{1}\cdot\bn\right)  \left(  \bS_{2}\cdot\bn\right)
+\bS_{1}\left(  \bS_{2}\cdot\bn\right)  +\bS%
_{2}\left(  \bS_{1}\cdot\bn\right)  \right]  ,\label{as1s2lo}%
\end{equation}
and
\begin{equation}
\ba^{s^2}_{\rm cons}=\sum\limits_{a\neq b}\frac{3m_{b}}{2m_{a}%
}\frac{C_{ES^{2}}^{(a)}}{m\nu}\frac{1}{r^{4}}\left[  -\bn\,\bS_{a}%
^{2}+{5\bn}\left(  \bS_{a}\cdot\bn\right)  ^{2}%
-2\bS_{a}\left(  \bS_{a}\cdot\bn\right)  \right].\label{asquadlo}%
\end{equation}
We will use these expressions in what follows in order to include spin-spin effects that contribute through derivatives of spin-independent multipole moments, just as we did in~\cite{paper1}.

\subsection{Nonconservative dynamics}

As we discussed in~\cite{paper1}, following the in-in formalism \cite{inin1,inin2}, the nonconservative dynamics is described by an extended Routhian,
\begin{align}
 \label{Reff1}
	{\cal R}_{\rm eff} [ \bx_{a \pm} ,S^{\mu\nu}_{ a \pm} ] = {\cal R}^{\rm cons}_{\rm eff} [ \bx_{a(1)},S^{\mu\nu}_{a(1)}] - {\cal R}^{\rm cons}_{\rm eff} [ \bx_{a(2)},S^{\mu\nu}_{a(2)}] - {\cal R}^{\rm RR}_{\rm eff} [ \bx_{a \pm} ,S^{\mu\nu}_{ a \pm} ] .
\end{align} 
To the PN order we work in this paper, the dissipative term is given by \cite{paper1}
\begin{align}
\label{seffR} 
	{\cal R}^{\rm RR}_{\rm eff}[\br_\pm,S^{\mu\nu}_\pm] = -\frac{1}{5} \, I_{-}^{ij}(t)I_{+}^{ij(5)}(t)-\frac{16}{45}\, J_{-}^{ij}(t) J_{+}^{ij(5)}(t)\,,
\end{align}
in terms of the $\pm$-variables. 
To compute spin-spin effects at leading order, we do not require the temporal components of the spin tensor, which implies that we can concentrate on the spin $3$-vector. 
The acceleration describing radiation reaction is obtained from~\eqref{seffR} via
\begin{align}
	\ba_{\rm RR}^{i} = \frac{1}{m\nu} \left[  \frac{\partial {\cal R}^{\rm RR}_{\rm eff}(\br_\pm,\bv_\pm,\bS_\pm)}{\partial\br_{i-}(t)}-\frac{d}{dt}\left(  \frac{\partial {\cal R}^{\rm RR}_{\rm eff}(\br_{\pm},\bv_{\pm},\bS_\pm)}{\partial\bv_{i-}(t)}\right)  \right]  _{\rm PL}\,,
\label{eq:ER}
\end{align}
whereas for the spin dynamics we have
\begin{align}
\label{eq:SR}
	\dot \bS_{\rm RR} = \lbracket \bS_{ +} ,{\cal R}^{\rm RR}_{\rm eff}(\br,\bv,\bS_{\pm}) \rbracket_{\rm PL}\,,
\end{align}
where ``PL" stands for the physical limit described in~\cite{paper1}. 
Notice that, in deriving the equation of motion for the spin, we can use the standard physical variables of position and velocities since these are simply spectators in~\eqref{eq:SR}.  
The required generalized Poisson brackets to be used in~\eqref{eq:SR} are given by,
\begin{align}
\label{eq:spinAlgebra1}
	\lbracket  \bS_{+}^{i},\bS_{+}^{j} \rbracket   = {} & -\frac{1}{4}\epsilon^{ijk}\bS_{-}^{k} \,,\\
	\lbracket \bS_{-}^{i},\bS_{-}^{j} \rbracket   = {} & - \epsilon^{ijk}\bS_{-}^{k} \,,\nn \\
	\lbracket \bS_{+}^{i},\bS_{-}^{j} \rbracket  = {} & - \epsilon^{ijk}\bS_{+}^{k}\,.\nn
\end{align}
See~\cite{paper1} for more details.

\section{Spin-Spin radiation reaction dynamics at 4.5PN order}

\subsection{Source multipoles}\label{secm}

The multipole moments to compute radiation-reaction due to spin are given in \cite{paper1} in terms of the $\pm$-variables. 
In the spin-spin sector to the desired order we need
\begin{align}
	I_{(0)-}^{ij} = {} & m\nu\left[  \br_{+}^{i}\br_{-}^{j}+\br_{-}^{i}\br_{+}^{j}\right]  _{\rm TF}\, , \\
	I_{(0)+}^{ij} ={} & m\nu\left[  \br_{+}^{i}\br_{+}^{j}\right]  _{\rm TF}\, , \\
	J_{S(0)-}^{ij} = {} & -\frac{3\nu}{2}\left[  \bSigma^{i}_+\br_{-}^j +  \bSigma^{i}_{-}\br_{+}^j \right]  _{\rm STF} \label{Jmin}\,, \\
	J_{S(0)+}^{ij} = {} & -\frac{3\nu}{2}\left[  \bSigma^{i}_{+}\br_{+}^{j}\right]  _{\rm STF}\,,
\end{align}
where ``(S)TF'' stands for (symmetric) trace-free and  
\begin{equation}
\bSigma = \frac{m}{m_2} \bS_2 - \frac{m}{m_1}\bS_1.
\end{equation} 
In addition, we also require the contribution to the mass quadrupole from finite size effects~\cite{srad,amps}
\begin{align}
I_{S^2(0)+}^{ij}=-&\sum_ a \frac{C^{(a)}_{ES^2}}{m_a}\left[\bS^i_{a+} \bS^j_{a+}\right]  _{TF}\,,\\
I_{S^2(0)-}^{ij}=-&\sum_ a \frac{C^{(a)}_{ES^2}}{m_a}\left[\bS^i_{a+} \bS^j_{a-} + \bS^i_{a-}\bS^j_{a+}\right]  _{TF}\,.\label{finiteQ}
\end{align}

\subsection{Acceleration}
\label{sec:acc}

We next derive the radiation-reaction accelerations. As in~\cite{paper1,chadbr2} we split the contribution into pieces. 
In our case, we have two contributions only from the current-quadrupole (``cq'') and a piece from order-reduction (``red''), which are given by
\begin{equation}
\ba_{\rm RR (cq)}^{m}= \frac{8}{15m}\left[
\boldsymbol{\Sigma}^{i}\delta^{jm}\right]  \frac{d^{5}J_{S(0)}^{ij}}{dt^{5}}\,,
\end{equation}
and
\beq
\label{eq:accRRred1}
\ba_{\rm RR (red)}^{m}=-\frac{2}{5}\left[  \br^{i}\delta^{jm}\right]
\left[  \frac{d^{5}I_{(0)}^{ij}}{dt^{5}}\right]_{SS},
\eeq
where the leading order spin-spin equations of motion in~\eqref{as1s2lo} and~\eqref{asquadlo} are used in the leading order mass quadrupole. 
In principle, we should add a contribution from \eqref{finiteQ} to the radiation-reaction force. 
However, since the time derivative of the spin vector introduces an extra $v^2$ and there is no explicit dependence on $\br$ or $\bv$ in \eqref{finiteQ}, such a contribution vanishes at leading order.

After some algebra, we obtain
\begin{align}
\label{eq:accRRcq1}
\ba_{\rm RR (cq)}  = {} &\frac{2\nu}{r^{7}} \bigg\{  3\left(  {\bSigma} \cdot \boldsymbol{\Sigma}\right)  \left[  -r\dot{r}\br\left(  2\frac{m}{r}-3v^{2}+7\dot{r}^{2}\right)  + r^{2}\bv\left(  \frac{8}{15}\frac
{m}{r}-\frac{3}{5}v^{2}+3\dot{r}^{2}\right)  \right]  \nn \\
	& \qquad   + r \boldsymbol{\Sigma}\left[  -\dot{r}\left(  \boldsymbol{\Sigma} \cdot \br \right) \left(  2\frac{m}{r}-3v^{2}+7\dot{r}^{2}\right) + r\left(  \boldsymbol{\Sigma}\cdot\bv\right)  \left(  \frac{8}{15}\frac{m}{r}-\frac{3}{5}v^{2}+3\dot{r}^{2}\right)  \right]  \bigg\} ,
\end{align}
for the current-quadrupole term whereas the reduced part in~\eqref{eq:accRRred1} can be split into two pieces:  the ``structureless'' contribution,
\begin{align}
\ba^{s_1s_2}_{\rm RR(red)} = {} & \frac{2}{5r^{7}} \bigg\{  \br \bigg[ r\dot{r}\left(  \bS_{1}\cdot\bS_{2}\right)  \left(  -112\frac{m}{r}+480v^{2}-980\dot{r}^{2}\right)  +\frac{\dot{r}}{r} \left( \bS_{1}\cdot\br\right)  \left(  \bS_{2} \cdot \br\right)  \left(  392\frac{m}{r}-3990v^{2}+10710\dot{r}^{2}\right) \nn  \\
	& \qquad  + \left(  \left(  \bS_{1}\cdot\br\right)  \left( \bS_{2}\cdot\bv\right)  + \left(  \bS_{1}\cdot
\bv\right)  \left(  \bS_{2}\cdot\br\right)  \right) \left(  -79\frac{m}{r}+555v^{2}-3465\dot{r}^{2}\right) + 960r\dot{r}\left( \bS_{1}\cdot\bv\right)  \left(  \bS_{2}\cdot \bv\right)  \bigg]  \nn \\
	&  + 6 \bv \bigg[  2r^{2}\left(  \bS_{1}\cdot\bS_{2}\right) \left(  8\frac{m}{r}-10v^{2}+45\dot{r}^{2}\right)  +\left(  \bS_{1}\cdot\br\right)  \left(  \bS_{2}\cdot\br\right) \left(  -51\frac{m}{r}+115v^{2}-735\dot{r}^{2}\right)  \nn \\
	& \qquad + 205r\dot{r}\left(  \left(  \bS_{1}\cdot\br\right) \left(  \bS_{2}\cdot\bv\right)  +\left(  \bS_{1} \cdot\bv\right)  \left(  \bS_{2}\cdot\br\right)  \right) - 40r^{2}\left(  \bS_{1}\cdot\bv\right)  \left(  \bS_{2}\cdot\bv\right)  \bigg]  \nn \\
	&   + 3r\bS_{1}\left[  \dot{r}\left(  \bS_{2}\cdot\br \right)  \left(  17\frac{m}{r}+80v^{2}-140\dot{r}^{2}\right)  +r\left( \bS_{2}\cdot\bv\right)  \left(  3\frac{m}{r}-15v^{2}+55\dot{r}^{2}\right)  \right]  \nn \\
	&  +3r\bS_{2}\left[  \dot{r}\left(  \bS_{1}\cdot\br\right)  \left(  17\frac{m}{r}+80v^{2}-140\dot{r}^{2}\right) +r\left(  \bS_{1}\cdot\bv\right)  \left(  3\frac{m}{r} -15v^{2}+55\dot{r}^{2}\right)  \right]  \bigg\}\,,
\end{align}
and a finite-size contribution given by
\begin{align}
\label{eq:accRRredFS1}
\ba_{\rm RR(red)}^{\rm FS}  = {} & \frac{2}{5r^{7}}\sum\limits_{a\neq b}\frac{C_{ES^{2}}^{(a)}m_{b}}{m_{a}} \bigg\{  \br \bigg[  -2r\dot{r}\bS_{a}^{2}\left(  28\frac{m}{r}-120v^{2}+245\dot{r}^{2}\right)+ 7\frac{\dot{r}}{r}\left(  \bS_{a}\cdot\br\right)^{2} \left( 28 \frac{m}{r}-285v^{2}+765\dot{r}^{2}\right)  \nn  \\
	& \qquad + \left(  \bS_{a}\cdot\br\right)  \left( \bS_{a}\cdot\bv\right)  \left(  -79\frac{m}{r}+555v^{2} - 3465\dot{r}^{2}\right)  +480r\dot{r}\left(  \bS_{a}\cdot \bv\right)  ^{2} \bigg]  \nn \\
	& + 3 \bv \bigg[  2r^{2}\bS_{a}^{2}\left(  8\frac{m}{r} - 10v^{2}+45\dot{r}^{2}\right)  +\left(  \bS_{a}\cdot\br\right)^{2}\left(  -51\frac{m}{r}+115v^{2}-735\dot{r}^{2}\right)  \nonumber\\
	&  \qquad + 410r\dot{r}\left(  \bS_{a}\cdot\br\right)  \left( \bS_{a}\cdot\bv\right)  -40r^{2}\left(  \bS_{a}
\cdot\bv\right)  ^{2} \bigg]  \nonumber\\
	&  +3r\bS_{a} \bigg[  \dot{r}\left(  \bS_{a} \cdot \br \right)  \left(  17\frac{m}{r}+80v^{2}-140\dot{r}^{2}\right) + r\left(  \bS_{a}\cdot\bv\right)  \left(  3\frac{m}{r} - 15v^{2}+55\dot{r}^{2}\right)  \bigg]  \bigg\}  .
\end{align}
To our knowledge, this is the first time the above expression has been computed.

The total spin 4.5PN acceleration is the sum of the contributions in~\eqref{eq:accRRcq1}--\eqref{eq:accRRredFS1}. 
It is also instructive to decompose the final expression into two pieces,
\begin{align}
\ba^{s_1s_2}_{\rm RR}  = {} & \frac{2}{r^{7}} \bigg\{  \br \bigg[ - r\dot{r}\left(  \bS_{1}\cdot\bS_{2}\right)  \left(  \frac{52}{5}\frac{m}{r}-78v^{2}+154\dot{r}^{2}\right)  +\frac{\dot{r}}{r}\left( \bS_{1}\cdot\br\right)  \left(  \bS_{2} \cdot \br\right)  \left(  \frac{392}{5}\frac{m}{r}-798v^{2}+2142\dot{r}^{2}\right) \label{ars1s2} \nn \\
	& \qquad  + \left(  \left(  \bS_{1}\cdot\br\right)  \left( \bS_{2}\cdot\bv\right)  +\left(  \bS_{1}\cdot
\bv\right)  \left(  \bS_{2}\cdot\br\right)  \right) \left(  -\frac{79}{5}\frac{m}{r}+111v^{2}-693\dot{r}^{2}\right)  +192r\dot{r}\left(  \bS_{1}\cdot\bv\right)  \left(  \bS_{2} \cdot \bv\right)  \bigg]  \nn \\
	&  + 2\bv \bigg[  r^{2}\left(  \bS_{1}\cdot\bS_{2}\right) \left(  8\frac{m}{r}-\frac{51}{5}v^{2}+45\dot{r}^{2}\right)  +3\left( \bS_{1}\cdot\br\right)  \left(  \bS_{2}\cdot \br\right)  \left(  -\frac{51}{5}\frac{m}{r}+23v^{2}-147\dot{r}^{2}\right)   \nn \\
	&  \qquad +123\, r\dot{r}\left(  \left(  \bS_{1}\cdot\br\right) \left(  \bS_{2}\cdot\bv\right)  +\left(  \bS_{1} \cdot \bv \right)  \left(  \bS_{2}\cdot\br\right)  \right) - 24r^{2}\left(  \bS_{1}\cdot\bv\right)  \left(  \bS_{2}\cdot\bv\right)  \bigg]  \nn \\
	&  + r\bS_{1} \bigg[  \dot{r}\left(  \bS_{2}\cdot\br \right)  \left(  \frac{61}{5}\frac{m}{r}+45v^{2}-77\dot{r}^{2}\right) + r\left(  \bS_{2}\cdot\bv\right)  \left(  \frac{19}{15}\frac{m}{r}-\frac{42}{5}v^{2}+30\dot{r}^{2}\right)  \bigg]  \nn \\
	&   + r\bS_{2} \bigg[  \dot{r}\left(  \bS_{1} \cdot \br\right)  \left(  \frac{61}{5}\frac{m}{r}+45v^{2}-77\dot{r}^{2} \right)  + r\left(  \bS_{1}\cdot\bv\right)  \left(  \frac{19}{15}\frac{m}{r}-\frac{42}{5}v^{2}+30\dot{r}^{2}\right)  \bigg]  \bigg\}  ,
\end{align}
for the ${\cal O}(\bS_1\bS_2)$ terms whereas the ${\cal O}(\bS_a^2)$ contributions are given by
\begin{align}
\label{arss2}
\ba_{\rm RR}^{s^2} = {} &\frac{2}{r^{7}} \bigg\{  \left( \frac{m_{2}}{m_{1}}\bS_{1}^{2}+\frac{m_{1}}{m_{2}}\bS_{2}^{2}\right)  \left[  -3r\dot{r}\br\left(  2\frac{m}{r}-3v^{2}+7\dot{r}^{2}\right)  +r\bv\left(  \frac{8}{5}\frac{m}{r}-\frac{9}{5}v^{2}+9\dot{r}^{2} \right) \right] \nn  \\
	& \qquad + \frac{m^{2}\nu}{m_{1}^{2}}r\bS_{1}\left[  -\dot{r}\left( \bS_{1}\cdot\br\right)  \left(  2\frac{m}{r}-3v^{2}-7\dot{r}^{2}\right)  +r\left(  \bS_{1}\cdot\bv\right)  \left(  \frac{8}{15}\frac{m}{r}-\frac{3}{5}v^{2}+3\dot{r}^{2}\right)  \right]  \nn \\
	& \qquad  +\frac{m^{2}\nu}{m_{2}^{2}}r\bS_{2}\left[  -\dot{r}\left( \bS_{2}\cdot\br\right)  \left(  2\frac{m}{r}-3v^{2}-7\dot{r}^{2}\right)  +r\left(  \bS_{2}\cdot\bv\right)  \left(  \frac{8}{15}\frac{m}{r}-\frac{3}{5}v^{2}+3\dot{r}^{2}\right)  \right]  \bigg\} \nn \\
	&  + \frac{2}{5r^{7}}\sum\limits_{a\neq b}\frac{C_{ES^{2}}^{(a)}m_{b}}{m_{a}} \bigg\{  \br \bigg[  -2r\dot{r}\bS_{a}^{2}\left(  28\frac{m}{r}-120v^{2}+245\dot{r}^{2}\right)  +7\frac{\dot{r}}{r}\left(  \bS_{a}\cdot\br\right)  ^{2}\left(  28\frac{m}{r}-285v^{2}+765\dot{r}^{2}\right)   \nn \\
	& \qquad \qquad +\left(  \bS_{a}\cdot\br\right)  \left( \bS_{a}\cdot\bv\right)  \left(  -79\frac{m}{r}+555v^{2} - 3465\dot{r}^{2}\right)  +480r\dot{r}\left(  \bS_{a}\cdot \bv\right)  ^{2} \bigg]  \nn \\
	& \qquad + 3\bv \bigg[  2r^{2}\bS_{a}^{2}\left(  8\frac{m}{r} - 10v^{2}+45\dot{r}^{2}\right)  +\left(  \bS_{a}\cdot\br\right)^{2}\left(  -51\frac{m}{r}+115v^{2}-735\dot{r}^{2}\right) \nn \\
	& \qquad \qquad + 410r\dot{r}\left(  \bS_{a}\cdot\br\right)  \left( \bS_{a}\cdot\bv\right)  -40r^{2}\left(  \bS_{a} \cdot \bv \right)^{2} \bigg]  \nn \\
	& \qquad  + 3r\bS_{a}\left[  \dot{r}\left(  \bS_{a} \cdot \br \right)  \left(  17\frac{m}{r}+80v^{2}-140\dot{r}^{2}\right) + r\left(  \bS_{a}\cdot\bv\right)  \left(  3\frac{m}{r} -15v^{2}+55\dot{r}^{2}\right)  \right]  \bigg\} .
\end{align}

\subsection{Spin evolution}

For the spin evolution we use \eqref{eq:SR}, \eqref{eq:spinAlgebra1}, and \eqref{seffR} to find
\begin{align}
{\dot{{\boldsymbol{S}}}}_{1\, \rm RR}^{m} = {} &-\frac{8\nu}{15}\frac{m}{m_{1}}\left[
J_{S(0)+}^{ij(5)}\br^{j}\lbracket {{\boldsymbol{S}}}_{1+}%
^{m},{{\boldsymbol{S}}}_{1-}^{i}\rbracket  \right]_{\rm PL} -\frac{2}{5}\frac{C_{ES^{2}}^{(1)}
}{m_{1}}\left[  I_{(0)+}^{ij(5)}{\bS}_{1+}^{i}\lbracket
{\bS}_{1+}^{m},{\bS}_{1-}^{j}\rbracket  \right]_{\rm PL}  \\
	= {} & \frac{8\nu}{15}\frac{m}{m_{1}}J_{S(0)}%
^{ij(5)}\epsilon^{mik}\br^{j}{{\boldsymbol{S}}}_{1}^{k}+\frac{2}{5}\frac{C_{ES^{2}%
}^{(1)}}{m_{1}}\epsilon^{mjk}\bS_{1}^{i}\bS_{1}^{k}%
I_{(0)}^{ij(5)}\,.\nn
\end{align}
Plugging the expression for the spin-independent mass- and current-quadrupole and order-reducing (time derivatives of) the acceleration using lower order PN equations of motion, we obtain:
\begin{align}
{\dot{{\boldsymbol{S}}}}_{1\, \rm RR} = {} &\frac{2m\nu}{15r^{6}} \bigg\{  6r^{2}\dot{r}\left(  {{\boldsymbol{S}}}_{1}\times{{\boldsymbol{S}}}_{2}\right)  \left( 11\frac{m}{r}-18v^{2}+30\dot{r}^{2}\right)    - 3r\left(  {{\boldsymbol{S}}}_{1}\times{{\boldsymbol{v}}}\right)  \left( \left(  {{\boldsymbol{S}}}_{2}\cdot{{\boldsymbol{r}}}\right)  -\frac{m_{2}}{m_{1}}\left(  {{\boldsymbol{S}}}_{1}\cdot{{\boldsymbol{r}}}\right)  \right) \left(  8\frac{m}{r}-9v^{2}+45\dot{r}^{2}\right)  \nn \\
	& \qquad +\left(  {{\boldsymbol{S}}}_{1}\times{{\boldsymbol{r}}}\right)  \bigg[ 15\dot{r}\left(  \left(  S_{2}\cdot{{\boldsymbol{r}}}\right)  -\frac{m_{2}}{m_{1}}\left(  {{\boldsymbol{S}}}_{1}\cdot{{\boldsymbol{r}}}\right)  \right) \left(  2\frac{m}{r}-3v^{2}+7\dot{r}^{2}\right)   \nn \\
	& \qquad \qquad \qquad\quad  +   2r\left(  \left(  {{\boldsymbol{S}}}_{2}\cdot {{\boldsymbol{v}}}\right)  -\frac{m_{2}}{m_{1}}\left(  {{\boldsymbol{S}}}_{1}\cdot{{\boldsymbol{v}}}\right)  \right)  \left(  8\frac{m}{r} - 9v^{2}+45\dot{r}^{2}\right)  \bigg]  \bigg\}   \nn \\
	&  -\frac{4C_{ES^{2}}^{(1)}m_{2}}{5r^{5}} \bigg\{  4\left(  \frac{m}{r} - 3v^{2}+15\dot{r}^{2}\right)  \left[  \left(  {\br}\times{\bS}_{1}\right)  \left(  {\bS}_{1}\cdot{\bv}\right)  +\left( {\bv}\times{\bS}_{1}\right)  \left(  {\bS}_{1}\cdot {\br}\right)  \right]   \nn \\
	&  \qquad\qquad\qquad +15\frac{\dot{r}}{r}\left(  {\br}\times{\bS}_{1}\right) \left(  {\bS}_{1}\cdot{\br}\right)  \left(  3v^{2}-7\dot{r}^{2}\right)  -30r\dot{r}\left(  {\bv}\times{\bS}_{1}\right)  \left( {\bS}_{1}\cdot{\bv}\right)  \bigg\}  .
\end{align}

It turns out this evolution equation may be written in a much more compact form, namely,
\begin{equation}
\left(\frac{d \tilde\bS_{1}}{dt}\right)_{\rm RR} {\hskip-0.1in} =  {\bm \Omega}^{RR}_1 \times \tilde \bS_1 \,,
\end{equation}
where 
\beq
{\bm \Omega}^{RR}_1 \equiv \frac{m}{3r^{6}} \left(\tilde \bS_2 - \frac{m_2}{m_1}\tilde \bS_1\right)\times \bL\,,
\eeq
can be identified as a precession frequency induced by radiation reaction effects on the spin's evolution, and
\begin{align}
\tilde \bS_1 = {} & \bS_1 +   \frac{m\nu}{15r^{5}}\Bigg\{  3r^{2}\left( {{\boldsymbol{S}}}_{1}\times{{\boldsymbol{S}}}_{2}\right)  \left(  3\frac{m}{r}-8v^{2}+24\dot{r}^{2}\right)  \nn  \\
	& \qquad - \frac{48r\dot{r}}{m\nu}\left[  {{{\boldsymbol{L}}}}\left(  {{\boldsymbol{S}}}_{1}\cdot{{\boldsymbol{S}}}_{2}\right)  -{{\boldsymbol{S}}}_{2} \left( {{{\boldsymbol{L}}}\cdot{\boldsymbol{S}}}_{1}\right)  +\frac{m_{2}}{m_{1}}\Big(  {{\boldsymbol{S}}}_{1}\left(  {{{\boldsymbol{L}}} \cdot {\boldsymbol{S}}}_{1}\right)  -{{{\boldsymbol{L}}}({\boldsymbol{S}}}_{1}\cdot \bS_1)\Big)  \right]  \nn \\
	& \qquad +3\left(  {{\boldsymbol{S}}}_{1}\times{{\boldsymbol{r}}}\right)  \left[ \left(  \left(  {{\boldsymbol{S}}}_{2}\cdot{{\boldsymbol{r}}}\right) - \frac{m_{2}}{m_{1}}\left(  {{\boldsymbol{S}}}_{1}\cdot {{\boldsymbol{r}}}\right)  \right)  \left(  \frac{m}{r}-2v^{2}+10\dot{r}^{2}\right)  - 2r\dot{r}\left(  \left(  {{\boldsymbol{S}}}_{2}\cdot{{\boldsymbol{v}}}\right) - \frac{m_{2}}{m_{1}}\left(  {{\boldsymbol{S}}}_{1}\cdot{{\boldsymbol{v}}}\right)  \right)  \right]  \nn \\
	& \qquad - 2r^{2}\left(  {{\boldsymbol{S}}}_{1}\times{{\boldsymbol{v}}}\right) \left(  \left(  {{\boldsymbol{S}}}_{2}\cdot{{\boldsymbol{v}}} \right)  -\frac{m_{2}}{m_{1}}\left(  {{\boldsymbol{S}}}_{1}\cdot {{\boldsymbol{v}}}\right)  \right) +\frac{2 m\nu}{5} \frac{C_{ES^{2}}^{(1)}}{m_{1}}\frac{d^{4}}{dt^{4}}\Big[ \left( {\bS}_{1}\cdot{\br}\right)  \left(  {\br}\times{\bS}_{1}\right)  \Big]   \Bigg\}  .
\end{align}
Notice that the spin equation describes precessing evolution with a constant norm. 
Furthermore, the effects due to the finite size contributions drop out of the final expression, turning into a total time derivative.  
Similar expressions and conclusions for the spin evolution of the second body can be found by interchanging the labels $1 \leftrightarrow 2$.

\section{Consistency test}

Here we perform a consistency test similar to the one presented in~\cite{paper1}. 
We first compute the radiated power in the far zone,
\begin{align}
\left(  \frac{dE}{dt}\right)_{SS}  = \left(  \frac{dE}{dt}\right) _{s_1s_2} + \left(  \frac{dE}{dt}\right) _{s^2} \,,
\end{align}
using the standard multipole formula. We find
\begin{align}
\left(  \frac{dE}{dt}\right) _{s_1s_2}  = {} &\frac{4m^{2}\nu}{15r^{8}%
} \big\{  -3r^{2}\left(  \bS_{1}\cdot\bS_{2}\right)  \left(
47v^{2}-55\dot{r}^{2}\right)  +3\left(  \bS_{1}\cdot\br\right)
\left(  \bS_{2}\cdot\br\right)  \left(  168v^{2}-269\dot{r}%
^{2}\right)  \nn \\
&  \qquad\quad+171\dot{r}\left(  \left(  \bS_{1}\cdot\br\right)
\left(  \bS_{2}\cdot\bv\right)  +\left(  \bS_{1}%
\cdot\bv\right)  \left(  \bS_{2}\cdot\br\right)  \right)
-71\left(  \bS_{1}\cdot\bv\right)  \left(  \bS_{2}%
\cdot\bv\right)  \big\}  ,
\end{align}
and
\begin{align}
\left(  \frac{dE}{dt}\right)_{s^2}  = {} & -\frac{2m^{4}\nu^{2}%
}{15r^{8}} \Bigg\{  3r^{2}\left(  \frac{\bS_{1}^{2}}{m_{1}^{2}}%
+\frac{\bS_{2}^{2}}{m_{2}^{2}}\right)  \left(  v^{2}+3\dot{r}%
^{2}\right)  +9\dot{r}^{2}\left(  \frac{\left(  \bS_{1}\cdot
\br\right)  ^{2}}{m_{1}^{2}}+\frac{\left(  \bS_{2}%
\cdot\br\right)  ^{2}}{m_{2}^{2}}\right)   \nn \\
	& \qquad\qquad\quad  -6r\dot{r}\left(  \frac{\left(  \bS_{1}\cdot\br%
\right)  \left(  \bS_{1}\cdot\bv\right)  }{m_{1}^{2}}%
+\frac{\left(  \bS_{2}\cdot\br\right)  \left(  \bS%
_{2}\cdot\bv\right)  }{m_{2}^{2}}\right)  +r^{2}\left(  \frac{\left(
\bS_{1}\cdot\bv\right)  ^{2}}{m_{1}^{2}}+\frac{\left(
\bS_{2}\cdot\bv\right)  ^{2}}{m_{2}^{2}}\right)  \Bigg\}
\nn \\
	&  -\frac{8}{5r^{8}}\sum\limits_{a\neq b}C_{ES^{2}}^{(a)}m_{b}^{2} \bigg\{
r^{2}\bS_{a}^{2}\left(  12v^{2}-13\dot{r}^{2}\right)  +2\left(
\bS_{a}\cdot\br\right)  ^{2}\left(  -21v^{2}+34\dot{r}%
^{2}\right)   \nonumber\\
	&  \qquad\qquad\qquad\qquad\quad  -29r\dot{r}\left(  \bS_{a}\cdot\br\right)  \left(
\bS_{a}\cdot\bv\right)  +6r^{2}\left(  \bS_{a}%
\cdot\bv\right)  ^{2} \bigg\} ,
\end{align}
for the ${\cal O}(\bS_1\bS_2)$ and ${\cal O}(\bS_a^2)$ contributions, respectively, in agreement with~\cite{kidder,Arun}. 
We then use the radiation-reaction acceleration computed in Section~\ref{sec:acc} to derive the instantaneous power from
\begin{align}
{\cal P}^{SS}_{\rm RR} \equiv \left( {\ba}^{s_1s_2}_{\rm RR}+\ba_{\rm RR}^{s^2}\right) \cdot\bv = {\cal P}^{s_1s_2}_{\rm RR}+{\cal P}^{s^2}_{\rm RR}\,.
\end{align}
Using \eqref{ars1s2} and \eqref{arss2} we find
\begin{align}
{\cal P}^{s_1s_2}_{\rm RR} = {} &\frac{4m\nu}{15r^{8}} \bigg\{  3r^{2}\left(
\bS_{1}\cdot\bS_{2}\right)  \left[  40mv^{2}+r\dot{r}^{2}\left(
-26\frac{m}{r}+420v^{2}\right)  -51rv^{4}-385r\dot{r}^{4}\right]  \nn \\
	& \qquad + 3\left(  \bS_{1}\cdot\br\right)  \left(  \bS_{2}%
\cdot\br\right)  \left[  -153mv^{2}+28r\dot{r}^{2}\left(  7\frac{m}%
{r}-150v^{2}\right)  +345rv^{4}+5355r\dot{r}^{4}\right]  \nonumber\\
	& \qquad - 3r^{2}\dot{r}\left[  \left(  \bS_{1}\cdot\br\right)  \left(
\bS_{2}\cdot\bv\right)  +\left(  \bS_{1}\cdot
\bv\right)  \left(  \bS_{2}\cdot\br\right)  \right]
\left(  9\frac{m}{r}-1005v^{2}+1925\dot{r}^{2}\right)  \nonumber\\
	& \qquad + r^{3}\left(  \bS_{1}\cdot\bv\right)  \left(
\bS_{2}\cdot\bv\right)  \left(  19\frac{m}{r}-486v^{2}%
+1890\dot{r}^{2}\right)  \bigg\}  
\end{align}
for the ${\cal O}(\bS_1\bS_2)$ contributions while the ${\cal O}(\bS_a^2)$ terms are given by
\begin{align}
{\cal P}^{s^2}_{\rm RR} = {} &-\frac{2m^{3}\nu^{2}}{15r^{8}} \Bigg\{
3r^{2}\left(  \frac{\bS_{1}^{2}}{m_{1}^{2}}+\frac{\bS_{2}^{2}%
}{m_{2}^{2}}\right)  \left[  -8mv^{2}+30r\dot{r}^{2}\left(  \frac{m}{r}%
-3v^{2}\right)  +9rv^{4}+105r\dot{r}^{4}\right]    \\
	& \qquad \qquad\quad +15r^{2}\dot{r}\left(  \frac{\left(  \bS_{1}\cdot\br\right)
\left(  \bS_{1}\cdot\bv\right)  }{m_{1}^{2}}+\frac{\left(
\bS_{2}\cdot\br\right)  \left(  \bS_{2}\cdot
\bv\right)  }{m_{2}^{2}}\right)  \left(  2\frac{m}{r}-3v^{2}+7\dot
{r}^{2}\right)  \nonumber\\
	& \qquad\qquad\quad  +r^{3}\left(  \frac{\left(  \bS_{1}\cdot\bv\right)
^{2}}{m_{1}^{2}}+\frac{\left(  \bS_{2}\cdot\bv\right)  ^{2}%
}{m_{2}^{2}}\right)  \left(  -8\frac{m}{r}+9v^{2}-45\dot{r}^{2}\right)
\Bigg\}  \nonumber\\
	&  -\frac{2}{5mr^{8}}\sum\limits_{a\neq b}C_{ES^{2}}^{(a)}m_{b}^{2} \bigg\{
2r^{2}\bS_{a}^{2}\left[  -24mv^{2}+r\dot{r}^{2}\left(  28\frac{m}%
{r}-255v^{2}\right)  +30rv^{4}+245r\dot{r}^{4}\right]   \nonumber\\
	& \qquad\qquad\qquad + \left(  \bS_{a}\cdot\br\right)  ^{2}\left[  153mv^{2}%
-28r\dot{r}^{2}\left(  7\frac{m}{r}-150v^{2}\right)  +345rv^{4}-5355r\dot
{r}^{4}\right]  \nonumber\\
	&  \qquad\qquad\qquad +r^{2}\dot{r}\left(  \bS_{a}\cdot\br\right)  \left(
\bS_{a}\cdot\bv\right)  \left(  28\frac{m}{r}-2025v^{2}+3885\dot
{r}^{2}\right)  -3r^{3}\left(  \bS_{a}\cdot\bv\right)
^{2}\left(  3\frac{m}{r}-55v^{2}+215\dot{r}^{2}\right)  \bigg\}  .\nonumber
\end{align}

The difference between the fluxes in the far zone and the instantaneous radiated power due to the radiation-reaction force reads
\begin{align}
\label{ep1} 
\left(  \frac{dE}{dt}\right)  _{s_1s_2} {\hskip-0.15in} - {\cal P}^{s_1s_2}_{\rm RR} = {} & \frac{4m\nu}{5r^{8}} \bigg\{  r^{2}\left(  \bS_{1}\cdot\bS%
_{2}\right)  \left[  -87mv^{2}+r\dot{r}^{2}\left(  81\frac{m}{r}%
-420v^{2}\right)  +51rv^{4}+385r\dot{r}^{4}\right]  \nn  \\
	& \qquad +\left(  \bS_{1}\cdot\br\right)  \left(  \bS_{2}%
\cdot\br\right)  \left[  321mv^{2}+r\dot{r}^{2}\left(  -465\frac{m}%
{r}+4200v^{2}\right)  -345rv^{4}-5355r\dot{r}^{4}\right]  \nonumber\\
	&\qquad  +r^{2}\dot{r}\left[  \left(  \bS_{1}\cdot\br\right)  \left(
\bS_{2}\cdot\bv\right)  +\left(  \bS_{1}\cdot
\bv\right)  \left(  \bS_{2}\cdot\br\right)  \right]
\left(  66\frac{m}{r}-1005v^{2}+1925\dot{r}^{2}\right)  \nonumber\\
	&  \qquad +6r^{3}\left(  \bS_{1}\cdot\bv\right)  \left(
\bS_{2}\cdot\bv\right)  \left(  -5\frac{m}{r}+27v^{2}-105\dot
{r}^{2}\right)  \bigg\} ,
\end{align}
and
\begin{align}
\label{ep2}
\left(  \frac{dE}{dt}\right) _{s^2} {\hskip-0.1in} - {\cal P}^{s^2}_{\rm RR} = {} &-\frac{2m^{3}\nu^{2}}{5r^{8}}\bigg\{  -3r^{2}\left(
\frac{\bS_{1}^{2}}{m_{1}^{2}}+\frac{\bS_{2}^{2}}{m_{2}^{2}%
}\right)  \left[  -3mv^{2}+r\dot{r}^{2}\left(  9\frac{m}{r}-30v^{2}\right)
+3rv^{4}+35r\dot{r}^{4}\right]  \nn \\
	& + 3m\dot{r}^{2}\left( \frac{\left(  \bS_{1}\cdot\br\right)^{2}}{m_{1}^{2}}+\frac{\left( \bS_{2}\cdot\br\right)^{2}}{m_{2}^{2}} \right) -r^{2}\dot{r}\left(  \frac{\left(  \bS_{1}\cdot\br\right)
\left(  \bS_{1}\cdot\bv\right)  }{m_{1}^{2}}+\frac{\left(
\bS_{2}\cdot\br\right)  \left(  \bS_{2}\cdot
\bv\right)  }{m_{2}^{2}}\right)  \left(  12\frac{m}{r}-15v^{2}%
+35\dot{r}^{2}\right)  \nn \\
	&   +3r^{3}\left(  \frac{\left(  \bS_{1}\cdot\bv\right)
^{2}}{m_{1}^{2}}+\frac{\left(  \bS_{2}\cdot\bv\right)  ^{2}%
}{m_{2}^{2}}\right)  \left(  \frac{m}{r}-v^{2}+5\dot{r}^{2}\right)  \bigg\}
\nonumber\\
	&  -\frac{2}{5mr^{8}}\sum\limits_{a\neq b}C_{ES^{2}}^{(a)}m_{b}^{2} \bigg\{
-2r^{2}\bS_{a}^{2}\left[  -48mv^{2}+3r\dot{r}^{2}\left(  18\frac{m}%
{r}-85v^{2}\right)  +30rv^{4}+245r\dot{r}^{4}\right]   \nn \\
	&  + 3\left(  \bS_{a}\cdot\br\right)  ^{2}\left[  -107mv^{2}%
+4r\dot{r}^{2}\left(  39\frac{m}{r}-350v^{2}\right)  +115rv^{4}+1785r\dot
{r}^{4}\right]  \nn \\
	&   +3r^{2}\dot{r}\left(  \bS_{a}\cdot\br\right)  \left(
\bS_{a}\cdot\bv\right)  \left(  -48\frac{m}{r}+675v^{2}%
-1295\dot{r}^{2}\right)  +3r^{3}\left(  \bS_{a}\cdot\bv\right)
^{2}\left(  11\frac{m}{r}-55v^{2}+215\dot{r}^{2}\right)  \bigg\} .
\end{align}
Similarly to spin-orbit effects in \cite{paper1}, the right-hand side of \eqref{ep1} and \eqref{ep2} can be shown to be a total derivative.
As in \cite{paper1}, we can also introduce Schott terms: $\tilde E = E - E_{\cal S}$, with
\begin{align}
E_{\cal S} = {} &   - \frac{1}{5}\left[  \frac{d^{2}I_{(0)}^{ij}}{dt^{2}%
}\frac{d^{3}I_{(0)}^{ij}}{dt^{3}}-\frac{dI_{(0)}^{ij}}{dt}\frac{d^{4}%
I_{(0)}^{ij}}{dt^{4}}\right]_{SS}  +\frac{2\nu}{5}\left[  2 \Big(  \bS%
_{1}\cdot\bS_{2}\right)  \left(  \ba\cdot{\dot{\ba}%
-\bv\cdot\ddot{\ba}}\right)     \\
	 & \qquad  +    \frac{1}{3}\Big(  \left(  \bS_{1}\cdot
{\dot{\ba}}\right)  \left(  \bS_{2}\cdot\ba\right)
+\left(  \bS_{1}\cdot\ba\right)  \left(  \bS_{2}%
\cdot{\dot{\ba}}\right)  -\left(  \bS_{1}\cdot{\ddot{\ba}%
}\right)  \left(  \bS_{2}\cdot\bv\right)  -\left(
\bS_{1}\cdot\bv\right)  \left(  \bS_{2}\cdot
{\ddot{\ba}}\right)  \Big)  \Big] \nonumber\\
	& - \frac{2m^{2}%
\nu^{2}}{15} \bigg[  3\left(  \frac{\bS_{1}^{2}}{m_{1}^{2}}%
+\frac{\bS_{2}^{2}}{m_{2}^{2}}\right)  \left(  \ba%
\cdot\dot{\ba}-\bv\cdot\ddot{\ba}\right)  +\frac{\left(  \bS_{1}\cdot\ba\right)
\left(  \bS_{1}\cdot\dot{\ba}\right)  }{m_{1}^{2}}+\frac{\left(
\bS_{2}\cdot\ba\right)  \left(  \bS_{2}\cdot\dot{\ba}\right)  }{m_{2}^{2}} \nn \\
	&  \qquad  -\frac{\left(  \bS_{1}%
\cdot\bv\right)  \left(  \bS_{1}\cdot\ddot{\ba}\right)
}{m_{1}^{2}}-\frac{\left(  \bS_{2}\cdot\bv\right)  \left(
\bS_{2}\cdot\ddot{\ba}\right)  }{m_{2}^{2}} \bigg] ,
\end{align}
where the explicitly written acceleration terms are given by the Newtonian expression, $\ba \to - (m/r^3) \br$, while the leading order spin-spin equations of motion~\eqref{as1s2lo} and~\eqref{asquadlo} are used for the evaluation of the derivatives in the mass-quadrupole terms. 
Hence,  
\beq
\left(\frac{d\tilde E}{dt}\right)_{SS} {\hskip-0.1in} = {\cal P}_{\rm RR}^{SS}\,,
\eeq
as expected. 
This concludes the consistency test in the spin-spin sector.

\section{Conclusions}

In this paper we continued the study initiated in~\cite{paper1} of nonconservative effects in the dynamics of spinning compact binary systems within an EFT framework~\cite{review}. 
We extended the formalism in~\cite{chadprl, chadprl2} to incorporate finite size effects, and computed the spin-spin contributions to the acceleration and spin evolution to 4.5PN order from first principles, without resorting to balance equations. To our knowledge, the calculation of finite size effects in radiation reaction had not been carried out until now.  
As in \cite{paper1}, we performed a consistency test by showing that the power induced by the radiation-reaction force is equivalent to the total radiated emission in the far zone, up to Schott terms. Unlike what we found in \cite{paper1}, the spin precesses due to spin-spin radiation reaction effects at this order. 
Our results are consistent with the findings in \cite{Will2} but we went a step further by extending the computations to all spin squared terms, including finite-size effects, which was not the case in~\cite{Will1,Will2}. 
Our results, which are the first in a series of contributions at 4.5PN order that are yet to be computed, will contribute to the modeling of spinning binary inspirals and the construction of more accurate waveforms, aiding in the extraction of precise information about gravitational wave sources from recent and future observations.

\section{Acknowledgements}

N.T.M. is supported by the Brazilian Ministry of Education (CAPES Foundation). 
C.R.G. is supported by NSF grant PHY-1404569 to Caltech. 
A.K.L. is supported by the NSF grant PHY-1519175. 
R.A.P is supported by the Simons Foundation and S\~ao Paulo Research Foundation (FAPESP) Young Investigator Awards, grants 2014/25212-3 and 2014/10748-5. 
R.A.P. also thanks the theory group at DESY (Hamburg) for hospitality while this work was being completed.
Part of this research was performed at the Jet Propulsion Laboratory, California Institute of Technology, under a contract with the National Aeronautics and Space Administration.

\bibliography{RefSO}

\end{document}